\documentstyle{article}
\author{L.J. Boya$^{z}$,
H. Rosu$^{l}$,
A.J. Segu\'{\i}-Santonja$^{z}$,
F.J. Vila$^{z}$\\
$^{z}$  Departamento de F\'{\i}sica Te\'orica, Facultad de Ciencias,\\
Universidad de Zaragoza, 50009-Zaragoza, Spain\\
$^{l}$  Instituto de F\'{\i}sica, Universidad de Guanajuato,
37150-Le\'on, Mexico\\}
\title{{\bf Strictly isospectral supersymmetry
and Schroedinger general zero modes}}
\begin{document}
\maketitle

{\bf Summary.} - The connection between the strictly isospectral
construction in supersymmetric quantum mechanics and the
general zero mode solutions of the Schroedinger equation is
explained by introducing slightly generalized first-order
intertwining operators. We also present a multiple-parameter generalization
of the strictly isospectral construction in the same perspective.

\bigskip


\bigskip

Nuovo Cimento B 113, 409-414 (March 1998) [quant-ph/9711059]

\vspace{4cm}


e-mail: luisjo@posta.unizar.es

e-mail: rosu@ifug3.ugto.mx

e-mail: segui@posta.unizar.es

e-mail: vila@posta.unizar.es

\newpage
{\bf 1}. - Consider the second order, one-dimensional, differential operator
of Schroedinger type
\begin{equation} \label{1}
H_{-}=-D^2+V_{-}(x),
\end{equation}
where $D=\frac{d}{dx}$ and $V_{-}(x)$ is a ``bosonic" type potential.
We are interested in solving
the eigenvalue equation associated with $H_{-}$ by means of factoring it
as the product of two first order operators.
If we consider the ground state case, and moreover take the ground state
energy as zero, we get the second order linear differential equation
$H_{-}\Phi _{-}=0$.
The general solution of such a linear equation will be
a linear composition of the two linearly independent particular solutions,
usually known as zero modes in the literature.
We shall suppose that at least one of the particular solutions,
say $u$, is nodeless and normalizable
(a true zero mode), a case denoted as unbroken supersymmetry \cite{w}.
If $u$ is supposed to be known, it allows us to
factorize $H_{-}$ in the following way \cite{w}. Defining $W=-{\rm ln} u$ as
the bosonic superpotential one gets
\begin{equation} \label{2}
H_{-}=-D^2+V_{-}(x)=A^{\dagger} A=(-D+W')(D+W'),
\end{equation}
where $W'=\frac{dW}{dx}$ and $V_{-}(x)=-W''+W'^{2}=\frac{u''}{u}$.
The true zero mode $u$ is annihilated by the $A$ operator $Au=0$.

We can construct a second linearly independent particular 
solution $v_{-}$ from the first one $u$, namely \cite{a}
\begin{equation} \label{3}
v_{-}=u \int ^{x}{\frac{1}{u^2}}.
\end{equation}
This solution is not annihilated by the $A$ operator, $Av_{-} \neq 0$ but
$A^{\dagger}$ acting on $Av_{-}=\frac{1}{u}$ is zero, since $v_{-}$ is a zero
mode of $H_{-}$.

The general solution annihilated by $H_{-}$ can be constructed as 
a linear combination of $u$ and $v_{-}$ as
$\Phi_{-}=au+bv_{-}=b(\frac{a}{b} u +v_{-})\propto(\lambda _{s} u +v_{-})$.
The range of variation of $\lambda _{s}$ can be extended such that
$\lambda _{s}\in (-\infty, \infty)$; $\lambda _{s}=0$ selects
the $v_{-}$ solution and $\lambda _{s}
\rightarrow \pm\infty$ the $u$ one. If we factorize $H_{-}$ by the
usual procedure but now using the general solution 
$\Phi_{-}(\lambda _{s})$ to construct a $\lambda _{s}$-dependent
superpotential 
$W(\lambda _{s})=-{\rm ln} \Phi _{-}(\lambda _{s})$, the factoring operators
$A$ and $A^{\dagger}$ will also depend on $\lambda _{s}$, but nevertheless
the $\lambda _{s}$ dependence disappears in the final result
and one ends up with the original potential $V_{-}$
\begin{equation} \label{4}
H_{-}(\lambda _{s})=(-D+W'(\lambda _{s}))(D+W'(\lambda _{s}))=
A^{\dagger}(\lambda _{s})
A(\lambda _{s})=-D^{2}+V_{-}(x).
\end{equation}

The supersymmetric ``fermionic" partner of (\ref{1}) is constructed by
interchanging the order of the factors \cite{w}
\begin{equation} \label{5}
H_{+}=A A^{\dagger}= (D+W')(-D+W')=-D^{2}+W'^{2}+W''=-D^{2}+V_{+}(x).
\end{equation}
The ``fermionic" potential differs from the ``bosonic" one by
$-2W''$, and, as is well known, both operators $H_{+}$ and $H_{-}$
have the same spectrum apart from the ground state which disappears
in $H_{+}$; as a matter of fact, one can write down
\begin{eqnarray} \label{6}
\lefteqn{ H_{-}|n>= E_{n}|n> \Rightarrow A^{\dagger} A|n>=E_{n}|n>
\Rightarrow A(A^{\dagger} A|n>)=A(E_{n}|n>) {} }
	\nonumber\\
& & {} \Rightarrow AA^{\dagger}(A|n>)=E_{n}(A|n>)
\Rightarrow H_{+}|n>= E_{n}|n>,
\end{eqnarray}
and for the ground state, $A|0>=0$.

One particular solution of the differential equation $H_{+} \Phi _{+} =0$ is
$\frac{1}{u}$, which, as said before, is annihilated by $A^{\dagger}$;
we can construct the second linearly independent solution by the same
procedure as for $H_{-}$
\begin{equation} \label{7}
v_{+}=\frac{1}{u} \int ^{x}{u^2},
\end{equation}
and the general solution of $H_{+}\Phi _{+}=0$ reads
\begin{equation} \label{8}
\Phi_{+}(\lambda _{s})= \lambda _{s}\frac{1}{u} +v_{+} =
\frac{1}{u} \left(\lambda _{s} +\int ^{x}{u^2}\right)~,
\end{equation}
which we call a one-parameter fermionic zero mode.

Now again, if we use the general solution (\ref{8}) to factorize the
fermionic Hamiltonian we can do it in an infinite number of ways,
depending on the value of $\lambda _{s}$ but again as in the bosonic case
all factorizations are producing only the original potential.

The interesting property of $\Phi _{+}$ is that its inverse has the form of
the general wave function which is used to generate
one-parameter families of strictly isospectral potentials to the
original bosonic one in the supersymmetric approach \cite{m}. Thus
\begin{equation} \label{9}
\Psi _{\lambda _{s}}=\frac{1}{\Phi _{+}}=\frac{u}{\lambda _{s}+
\int ^{x}{u^2} }~,
\end{equation}
can be used to construct a family of Hamiltonians
\begin{equation} \label{10}
H(\lambda _{s})=A^{\dagger}(\lambda _{s})A(\lambda _{s})=
-D^2 +V_{-}(\lambda _{s}),
\end{equation}
where $A(\lambda _{s})=D+W(\lambda _{s})$ and $W(\lambda _{s})=-
{\rm ln}(\Psi _{\lambda _{s}})$.
These Hamiltonians are strictly isospectral (there is absolutely no difference
in their energy spectra, though there are differences in their wave functions)
since they can be constructed by means
of the general solution of the Riccati equation; the isospectral family
includes the original potential $V_{-}$ corresponding to the superposition
quotient
$\lambda _{s} \rightarrow \pm\infty$. We briefly present this method in the
following.   

\bigskip

{\bf 2}. - Consider the ``fermionic" Riccati equation (FRE)
$y^{'}=-y^2+f(x)$
with the known particular solution $y_{0}$ and let $y_1=w_1+y_{0}$ be the
general solution.
By substituting $y_{1}$ in FRE one gets the Bernoulli equation
$-w_{1}^{'}=w_{1}^2+(2y_{0})w_{1}$. Furthermore,
using $w_2=1/w_1$, the simple first-order linear differential equation
$w_{2}^{'}-(2y_{0})w_{2}-1=0$ is obtained, which can be solved
by employing the
integration factor $f=e^{-\int ^{x}2y_{0}}$, leading
to the solution
$w_2=e^{\int ^{x}2y_{0}}(\int ^{x}e^{-\int ^{z}2y_{0}}+\lambda _{r})$,
where $\lambda _{r}$ occurs as an integration constant. Thus, the
general solution of FRE is
\begin{equation} \label{11}
y_{1}=y_{0}+
\frac{e^{-\int ^{x}2y_{0}}}{\lambda _{r}+\int ^{x}
e^{-\int ^{z}2y_{0}}}\equiv
y_{0}+\frac{f}{\lambda _{r} +\int ^{x}(f)}~.
\end{equation}
One can see easily that the particular Riccati solution $y_{0}$ corresponds to
Witten's superpotential \cite{w} 
while the general Riccati solution
$y_1=y_{0}+\frac{f}{\lambda _{r}+\int ^{x}f}$ is of
Mielnik type \cite{m}. Also $u=f^{1/2}$ can be normalized being the
ground state wavefunction (the true zero mode), and $-2y_{0}^{'}$
($\equiv -2\frac{d^2}{dx^2}\ln f^{1/2}$)
is the Darboux transform part of the
Schroedinger potential. Moreover, one can see easily that the modes
\begin{equation} \label{12}
\Psi _{\lambda _{r}}=
\frac{f^{1/2}}{\lambda _{r}+\int^{x} f}=\frac{u}{\lambda _{r}+
\int^{x}u^2}
\end{equation}
can be normalized and therefore are the ground state wavefunctions
(true zero modes of the bosonic family)
corresponding to Mielnik's parametric superpotential. Also,
$-2y_{1}^{'}$ can be thought of as the general Darboux transform part
in the potential generating the bosonic strictly isospectral family, which
reads
\begin{equation} \label{13}
V_{\lambda _{r}}=V_{-}(x)-\frac{4uu^{'}}{\lambda _{r}+\int^{x}u^2}
+\frac{2u^4}{(\lambda _{r}+\int^{x}u^2)^2}~.
\end{equation}
This family of potentials can be seen as a continuous deformation of the
original potential, because this is included in the infinite limit of the
deforming parameter.
All these relationships are supplementary material to the core of
the ``entanglement" between Riccati and Schroedinger equations,
which has been recently emphasized by Haley \cite{h97}.

If now, the two $\lambda$ constants we have used are identified,
i.e., $\lambda _{r}
\equiv\lambda _{s}\equiv \lambda$, then of
course $\Psi _{\lambda _{r}}\equiv\Psi _{\lambda _{s}}$. Since
$\lambda _{s}$ is a superposition quotient while $\lambda _{r}$ is an
integration constant, and both can be fixed through boundary conditions,
we can see that the identification is sound. This also gives a clue on
the connection between Schroedinger and Riccati methods at the level
of their general solutions.

We have now to answer the important question why the factorization based
on $\Psi _{\lambda _{r}}$ gives a nontrivial and therefore a physically
relevant case and what mathematical feature lies behind (\ref{9}).
These issues are addressed in the following.

\bigskip

{\bf 3}. - The mathematical background which is needed here belongs to the
intertwining operator transformations studied mathematically by Moutard,
Darboux, Ince, Delsart, Lions, and others. For a detailed
analysis see \cite{d}. Intertwining has been used by Pursey in his studies
of the combined procedures for generating families of strictly
isospectral Hamiltonians \cite{p}, whereas Anderson \cite{an} applied matrix
intertwining relations to the Dirac equation showing that their structure
is described by an $N=4$ superalgebra.

Two operators $L_{0}$ and $L_{1}$ are said to be intertwined by an operator
$T$ if
\begin{equation} \label{14}
L_{1}T=TL_{0}~.
\end{equation}
If the eigenfunctions $\varphi _{0}$ of $L_0$ are known, then from the
intertwining relation one can show that the (unnormalized) eigenfunctions
of $L_1$ are given by $\varphi _{1}=T\varphi _{0}$.  The main problem in
the intertwining transformations is to construct the transformation
operator $T$. One-dimensional quantum mechanics is one of the simplest
examples of intertwining relations since Witten's transformation operator
$T_{qm}=T_1$ is just a first spatial derivative
plus a differentiable coordinate function (the superpotential) that should
be a logarithmic derivative of the true bosonic zero mode (if it exists),
but of course
higher-order transformation operators can be constructed without much
difficulty \cite{bs}.

Thus, within the realm of one-dimensional quantum mechanics, writing $T_1=
D-\frac{u^{'}}{u}$, where $u$ is a true bosonic zero mode, one can
infer that the
adjoint operator $T^{\dagger}_{1}=-D-\frac{u^{'}}{u}$ intertwines in
the opposite direction, taking solutions of $L_{1}$ to those of $L_{0}$
\begin{equation}  \label{15}
\varphi _{0}=T_{1}^{\dagger}\varphi _{1}~.
\end{equation}
In particular, within quantum mechanics, $L_0=H_{-}$ and $L_1=H_{+}$ and
although the true zero mode of $H_{-}$ is annihilated by $T_1$, the
corresponding (unnormalized) eigenfunction of $H_{+}$ can nevertheless
be obtained by
applying $T_1$ to the other independent zero energy solution of $H_{-}$.
It is now not very difficult to see that a slightly modified transformation
operator of the adjoint type
\begin{equation}  \label{16}
T^{-}_{\lambda}=-D+\Psi _{\lambda}^2-\frac{\Psi _{\lambda}^{'}}
{\Psi _{\lambda}}
\end{equation}
when applied on the general zero mode solution $\Phi _{+}$ of $H_{+}$
will produce
precisely the strictly isospectral family of bosonic zero modes
($T_{\lambda}^{-}\Phi _{+}=\Psi _{\lambda}$). On the other hand, the
corresponding transformation operator sending general bosonic zero modes
to general fermionic ones reads
\begin{equation}  \label{17}
T^{+}_{\lambda} =D+\Psi _{\lambda}^{-2}-\frac{\Psi _{\lambda} ^{'}}
{\Psi _{\lambda}}
\end{equation}

Moreover, one can start with one of the strictly isospectral bosonic
zero modes $\Psi _{\lambda _{1}}=\frac{u}{\lambda _{1}+\int^{x} u^2}=u_1$
(i.e., fixing $\lambda =\lambda _{1}$)
and repeat the strictly isospectral construction, or in other words, using a
new two-parameter (of which only the second one is a free parameter) fermionic
zero mode of a Hamiltonian $H_{++}$
\begin{equation}  \label{18}
\Phi _{+}(\lambda _{1},\lambda _{2})=\frac{1}{u_{1}}\left(\lambda _{2}
+ \int ^{x}u_{1}^{2}
\right)~,
\end{equation}
whose inverse $\Psi _{\lambda _1,\lambda _{2}}=\frac{u_1}{\lambda
+\int^{x} u_{1}^{2}}=u_2$, $\lambda _{2}\in (-\infty, \infty)$, is just
a two-parameter family of bosonic zero modes. The resulting
two-parameter family of strictly isospectral potentials will be
\begin{equation}  \label{19}
V_{\lambda _{1}, \lambda _{2}}=
V_{\lambda _{1}}(x)-\frac{4u_2u^{'}_{2}}{\lambda _{2} +\int^{x}u_{2}^2}
+\frac{2u_{2}^4}{(\lambda _{2} +\int^{x}u_{2}^2)^2}~,
\end{equation}
where $V_{\lambda _{1}}$ is given by (\ref{13}),
with $\lambda _{r}=\lambda _{1}$.
Generalized formulas for multiple-parameter cases can be easily
written down, as well as a multiple parameter form of the modified
transform operators.

As we mentioned, the general bosonic zero modes,
$u_1$, $u_2$, $u_3$,...$u_{i},...$,
can be normalized and therefore turned into true zero modes. Their
normalization constants are of the type
$N_{i}=\sqrt{\lambda _{i}(\lambda _{i}+1)}$, where $i=1,2,3,...$, whenever
one is performing the normalization at the previous $(i-1)$-parameter step
and considering $u_{0}=u$ as a true zero mode (i.e., $N_{0}=1$, and thus
already normalized to unity). At an arbitrary $i$-level,
the parameter dependence of $u_{i}$, when expressed in terms of
only $u_{0}$ is a rather complicated
type of denominator possessing nesting integrals, which physically may be
considered as a modulational factor endowed to the multiple-parameter, bosonic,
true zero mode through the mathematical procedure. The parametric
normalization deletes the interval $[-1,0]$ from the parameter space;
at the $-1$ limit, one can make a connection with the Abraham-Moses
isospectral technique \cite{am}, whereas at the $0$ limit the connection
can be done
with another isospectral construction developed by Pursey \cite{p}.
Moreover, since the strictly isospectral supersymmetry obviously may
introduce singularities in both wavefunction and potential, usually the
active authors in the field are discarding those values of the deformation
parameter for which those singularities are occuring. If
$|\lambda _{{\rm sing}}|\geq 1$ the excluded interval in the parameter space
is $[-\lambda _{{\rm sing}}, \lambda _{{\rm sing}}]$, and therefore the
connections with the Abraham-Moses and Pursey methods are lost in this case.
Usually, at least one of the limits is lost, as for example
in the harmonic oscillator case where the excluded interval
is $[-\frac{\sqrt{\pi}}{2},\frac{\sqrt{\pi}}{2}]$ implying the loss of the
Pursey limit.
Also, the strictly isospectral supersymmetric method (as well as the other
isospectral procedures) may be considered as
a way of allowing for some physical effects of the irregular Schroedinger
(vacuum) solutions in quantum mechanics, which is not so much of a nonsense
nowadays \cite{leo}.
Indeed, the wrong vacuum is wrong only because of its asymptotic
behaviour; in some cases, the strictly isospectral constructions
show that the asymptotic behaviour of the wrong vacuum can be tamed.


\bigskip

In conclusion, we presented in some detail the parameter dependence of the
zero energy sector of the
unbroken supersymmetric quantum mechanics, which in applications is
usually fixed through some sort of boundary conditions.

\bigskip
\bigskip

\newpage

{\bf Acknowledgments}
\bigskip

This work was partially supported by the research grants AEN96-1670 (CSIC)
and ERBCHRX-CT92-0035 and by the CONACyT project 4868-E9406.

\end{document}